\documentclass[a4paper]{article}
\pdfoutput=1
\usepackage[hscale=0.62,vscale=0.8]{geometry}

\usepackage{cite}
\usepackage{amsmath,amssymb,amsfonts}
\usepackage[hyphens]{url}
\usepackage[hyperindex,breaklinks]{hyperref}
\usepackage{cite}

\usepackage[T1]{fontenc}

\usepackage{algorithm}
\usepackage{syntax}
\usepackage{fancyvrb}

\usepackage[table]{xcolor}

\usepackage{microtype}
\usepackage{multicol}
\usepackage[footnotesize,caption=false]{subfig}

\usepackage{multirow}
\usepackage{mathpartir}

\usepackage{semantic}

\usepackage{tikz}
\usetikzlibrary{arrows, automata, positioning}

\providecommand{\keywords}[1]{\textbf{\textit{Keywords ---}} #1}

{\itshape}{\rmfamily}

{\itshape}{\rmfamily}

\newcommand{\emptySeq}{\langle \rangle}
\newcommand{\Seq}[1]{\langle #1 \rangle}

\newcommand{\lsti}[1]{\lstinline|#1|}
\newcommand{\tlsti}[1]{\text{\lstinline|#1|}}

\newcommand{\isr}{~\textit{\textbf{is}}~}
\newcommand{\cat}{^\frown}

\newcounter{rowcounter}
\makeatletter
\newcommand{\xRightarrow}[2][]{\ext@arrow 0359\Rightarrowfill@{#1}{#2}}
\makeatother

\newcommand{\spath}{\xrightarrow}
\newcommand{\Path}{\xRightarrow}

\newcommand{\tbf}[1]{\textbf{#1}}

\newcommand{\ttbf}[1]{\texttt{\textbf{#1}}}

\newcommand{\tit}[1]{\textnormal{\textit{#1}}}

\def\boogie{\afterassignment\xcode\let\tmp}

\def\xcode{\setbox0\hbox\bgroup\aftergroup\setcode
	\lstinline[language=BoogieInline]}

\def\setcode{\usebox0}

\def\solidity{\afterassignment\solxcode\let\tmp}

\def\solxcode{\setbox0\hbox\bgroup\aftergroup\solsetcode
	\lstinline[language=SolidityInline]}

\def\solsetcode{\usebox0}

\def\BibTeX{{\rm B\kern-.05em{\sc i\kern-.025em b}\kern-.08em
		T\kern-.1667em\lower.7ex\hbox{E}\kern-.125emX}}

\makeatletter\@addtoreset{case}{theorem}\makeatother

\definecolor{verylightgray}{rgb}{.97,.97,.97}

\usepackage{listings}

\lstdefinelanguage{Solidity}{
	keywords=[1]{anonymous, assembly, assert, balance, break, call, callcode, case, catch, class, constant, continue, constructor, contract, debugger, default, delegatecall, delete, do, else, emit, event, experimental, export, external, false, finally, for, function, gas, if, implements, import, in, indexed, instanceof, interface, internal, is, length, library, log0, log1, log2, log3, log4, memory, modifier, new, payable, pragma, private, protected, public, pure, push, require, return, returns, revert, selfdestruct, send, solidity, storage, struct, suicide, super, switch, then, this, throw, transfer, true, try, typeof, using, value, view, while, with, addmod, ecrecover, keccak256, mulmod, ripemd160, sha256, sha3}, 
	keywordstyle=[1]\color{blue}\bfseries,
	keywords=[2]{address, bool, byte, bytes, bytes1, bytes2, bytes3, bytes4, bytes5, bytes6, bytes7, bytes8, bytes9, bytes10, bytes11, bytes12, bytes13, bytes14, bytes15, bytes16, bytes17, bytes18, bytes19, bytes20, bytes21, bytes22, bytes23, bytes24, bytes25, bytes26, bytes27, bytes28, bytes29, bytes30, bytes31, bytes32, enum, int, int8, int16, int24, int32, int40, int48, int56, int64, int72, int80, int88, int96, int104, int112, int120, int128, int136, int144, int152, int160, int168, int176, int184, int192, int200, int208, int216, int224, int232, int240, int248, int256, mapping, string, uint, uint8, uint16, uint24, uint32, uint40, uint48, uint56, uint64, uint72, uint80, uint88, uint96, uint104, uint112, uint120, uint128, uint136, uint144, uint152, uint160, uint168, uint176, uint184, uint192, uint200, uint208, uint216, uint224, uint232, uint240, uint248, uint256, var, void, ether, finney, szabo, wei, days, hours, minutes, seconds, weeks, years},	
	keywordstyle=[2]\color{teal}\bfseries,
	keywords=[3]{block, blockhash, coinbase, difficulty, gaslimit, number, timestamp, msg, data, gas, sender, sig, value, now, tx, gasprice, origin},	
	keywordstyle=[3]\color{violet}\bfseries,
	identifierstyle=\color{black},
	sensitive=true,
	comment=[l]{//},
	morecomment=[s]{/*}{*/},
	commentstyle=\color{gray}\ttfamily,
	stringstyle=\color{red}\ttfamily,
	morestring=[b]',
	morestring=[b]",
}

\lstdefinelanguage{Boogie}{
	keywords={var,:,where,while,if,else,procedure,returns,call,assume,assert,:=,havoc,int,bool,record,enum}, 
	keywordstyle=\color{black}\bfseries,
	identifierstyle=\color{black},
	sensitive=true,
	comment=[l]{//},
	commentstyle=\color{gray}\ttfamily,
	stringstyle=\color{red}\ttfamily,
	mathescape=true,
}

\lstdefinelanguage{BoogieInline}[]{Boogie}{
	basicstyle=\small\ttfamily,
	mathescape=true,
}

\lstdefinelanguage{SolidityInline}[]{Solidity}{
	basicstyle=\small\ttfamily,
	mathescape=true,
}

\begin{document}

\title{Formalising and verifying smart contracts with Solidifier: a bounded model checker for Solidity}

\author{Pedro Antonino \\
	\small
	\small \textit{The Blockchouse Technology Limited} \\
		\small Oxford, UK\\
		\small pedro@tbtl.com
	\and
		A. W. Roscoe \\
		\small		\textit{The Blockchouse Technology Limited} \\
		\small \textit{University College Oxford Blockchain Research Center}\\
		\small \textit{Department of Computer Science, Oxford University} \\
		\small Oxford, UK\\
		\small awroscoe@gmail.com
}

\date{}

\maketitle

\begin{abstract}
	The exploitation of smart-contract vulnerabilities can have catastrophic consequences such as the loss of millions of pounds worth of crypto assets. Formal verification can be a useful tool in identifying vulnerabilities and proving that they have been fixed. In this paper, we present a formalisation of Solidity and the Ethereum blockchain using the Solid language and its blockchain; a Solid program is obtained by explicating/desugaring a Solidity program. We make some abstractions that over-approximate the way in which Solidity/Ethereum behave. Based on this formalisation, we create Solidifier: a bounded model checker for Solidity. It translates Solid into Boogie, an intermediate verification language, that is later verified using Corral, a bounded model checker for Boogie. Unlike much of the work in this area, we do not try to find specific behavioural/code patterns that might lead to vulnerabilities. Instead, we provide a tool to find errors/bad states, i.e. program states that do not conform with the intent of the developer. Such a bad state, be it a vulnerability or not, might be reached through the execution of specific known code patterns or through behaviours that have not been anticipated.
\end{abstract}


\section{Introduction}
Smart contracts provide a new paradigm for \emph{trusted} execution of code~\cite{Szabo97,EthereumWhite}. A \emph{smart contract} is a program that runs in a trusted abstract machine/execution environment and has access to trusted data; its behaviour typically involves managing some digital asset. To achieve these properties, such a program is typically run in the context of a blockchain~\cite{EthereumWhite,EthereumYellow, Fabric,Sawtooth,Androulaki18} but other technologies can be used too~\cite{Wustace19,Cheng19,Kalodner18}. It can also be seen as a way to extend the capabilities of a blockchain: the code of a smart contract can be used to encode some transaction validation logic that was not originally implemented in/enforced by the blockchain~\cite{Fabric,Androulaki18}. Being a program, it can have code vulnerabilities that expose the digital assets it manages. When exploited, these vulnerabilities can lead to catastrophic effects such as the loss of vast sums of money~\cite{Atzei17}. The ability to reliably write transaction logic using smart contracts, and the flexibility that comes with it, will play a key role in the dissemination and adoption of blockchains and in the realisation of all their expected/predicted impact in society~\cite{Iansiti17, Swan15,Crosby16}. Hence, the soundness of smart contracts is a critical element in the blockchain framework that needs to be addressed.

Formal methods have been used in many contexts to ensure that systems behave as expected~\cite{Clarke96,Baier08,Roscoe10,Burch92,Grumberg94,Hoare85,Mota01,Barnett05,Rakamaric14,Antonino19}. It is only natural, then, that formal approaches have been proposed for the specification and verification of smart contracts~\cite{Kalra18,Luu16,Hildenbrandt18,Hajdu19,Hajdu20,Wang18,Bhargavan16,Grishchenko18,Tsankov18,Liu18,Alt18,Nehai18}. Most of these approaches focus on the Ethereum Virtual Machine (EVM) bytecode~\cite{EthereumYellow}, the low-level language used by the Ethereum blockchain~\cite{EthereumWhite}, and on finding specific patterns of code that might lead to a vulnerability using symbolic execution or static analysis. In this paper, we propose a framework to analyse (in our case falsify) semantic properties of programs in Solidity~\cite{Solidity}, a high-level language used by Ethereum smart-contract developers. Like  \cite{Hajdu19,Wang18,Permenev20}, we advocate the analysis of user-specified semantic properties of contracts, since their violations can exhibit known but also \emph{unanticipated} behavioural patterns leading to a vulnerability. Furthermore, this can be used to analyse the functional correctness of smart contracts.

We propose a formalisation of a representative subset of Solidity using our \emph{Solid} language, which is an explicated/desugared/simplified version of Solidity. It is intended to complement Solidity's documentation~\cite{Solidity}. We also introduce \emph{Solidifier}: a bounded model checker for Solidity. It is based around a translation of Solid into Boogie~\cite{Leino08,Barnett05}, an \emph{intermediate verification language}, that is ultimately checked by Corral~\cite{Lal12}, a bounded model checker for Boogie. Solidifier has been designed to find bugs, i.e. behaviours falsifying properties, which should be then fixed by developers and re-verified. We propose a new way to capture Solidity's unorthodox memory model, which is a key challenge in faithfully capturing its semantics. Moreover, our evaluation seems to suggest that not only Solidifier (and our Boogie encoding) more faithfully captures Solidity's semantics than similar tools, but it does so with an arguably insignificant verification speed overhead. 



\noindent
\textit{Outline.} Section~\ref{sec:background} briefly introduces key concepts of blockchains and formal verification. In Section~\ref{sec:related}, we describe some previous works that propose frameworks to formally reason about smart contracts. Section~\ref{sec:Solid} describes Solid and its blockchain, mimicking the behaviour of Solidity and the Ethereum blockchain. We present Solidifier in Section~\ref{sec:bmc}, and our final remarks in Section~\ref{sec:conclusion}.  

\section{Background}
\label{sec:background}

We now give a brief introduction to blockchain and formal verification concepts relevant to understanding this work.

\subsection{Blockchains and smart contracts}

A blockchain is a decentralised system where all nodes have a consistent view over its database. The concept of a blockchain was introduced to prevent double spending in the Bitcoin network~\cite{Nakamoto08} but, since then, it has been generalised in many ways, culminating in the proposal of smart contracts as a flexible mechanism to extend its capabilities~\cite{EthereumWhite,Androulaki18,Sawtooth}. A blockchain is a transaction-based system designed with a fixed set of rules that defines its basic transaction logic. For instance, they define what types of transactions can be processed by the blockchain and what makes a transaction \emph{valid}. These rules are enforced by a consensus protocol executed by the blockchain nodes; reaching a consensus is, to a large extent, what prevents nodes from misbehaving and, consequently, what ensures the correct behaviour of a blockchain. The blockchain's behaviour emerges from the processing of transactions and the effects it has on the underlying database. Once the blockchain network is deployed these rules are fixed and cannot be modified. So, one way to make the behaviour of a blockchain more flexible is to allow computation-triggering transactions, which are valid if the underlying computation completes without raising errors. This mechanism is precisely what smart-contract-based blockchains implement. To allow such a transaction, a blockchain has to define: what computations can be created, namely, what is the language that can be used to create such computing function; how they are computed, i.e., what are the semantics of these computations; and what is the process by which a computation can be made available to be executed via a transaction.

The Ethereum blockchain (or just Ethereum, for short) is arguably the most widely-known and used smart-contract-based blockchain~\cite{EthereumWhite}. In Ethereum, smart contracts are ultimately executed by the EVM and are represented using its \emph{bytecode}. Two high-level languages Solidity~\cite{Solidity} and Vyper~\cite{Vyper} are generally used to create smart contracts that are later compiled into EVM bytecode. An Ethereum transaction can be used to \emph{deploy} a smart contract to a blockchain \emph{address}. Once deployed, smart contract functions can be invoked using a contract-execution transaction by specifying what address is to be called and what function needs to be executed. Smart contracts in this platform also have a persistent state that is also stored at this address in the blockchain.

\subsection{Formal verification and bounded model checking}

Formal-verification techniques use mathematical models to precisely reason about the behaviour of systems. Most of them formalise some language in a framework that allows some degree of automation, scalability and precision in proving properties for programs in this language. A bounded model checker systematically and symbolically explores the behaviour of a system up to a certain bound $k$ looking for violations of a given property; the bound typically represents the number of transitions allowed to be taken from some initial state. 

In this work, we use Corral~\cite{Lal12} to implement our own Solidity model checker. Corral is a \emph{reachability-modulo-theories verifier} for the intermediate verification language Boogie~\cite{Leino08,Barnett05}. Traditional bounded model checkers create a symbolic constraint corresponding to the behaviour of an input system unwound to the given pre-specified bound $k$. Corral, on the other hand, uses counterexample-guided abstraction refinement (CEGAR) to (possibly) simplify and incrementally search the state space up to $k$. This abstraction process leads to an intermediate over-approximation of the input program that might even be sufficient to \emph{prove} properties. Its intended use, however, is to find property violations.

\section{Related work}
\label{sec:related}
The semantics of the EVM has been formalised in traditional interactive theorem provers~\cite{Hirai17,Amani18,Park18}. The work in~\cite{Hildenbrandt18} formalises the behaviour of the EVM using the $\mathbb{K}$ framework~\cite{Rosu10}, whereas the authors of~\cite{Grishchenko18} use $\text{F}^{*}$~\cite{Swamy16} to this end. These frameworks are designed to prove properties of systems as opposed to falsifying them. They are very precise and scalable (as they can establish properties of a complex system in a reasonable time) but they have a low degree of automation and require a considerable amount of manual effort. They are as precise and scalable as the proof engineer that manually guides the proving effort.

SmartCheck~\cite{Tikhomirov18}, Oyente~\cite{Luu16}, EtherTrust~\cite{Grishchenko18a} and Securify~\cite{Tsankov18}, in a fully automatic way, examine an approximation of the behaviour of a smart contract to look for specific behavioural/code patterns that tend to lead to vulnerabilities. SmartCheck~\cite{Tikhomirov18} translates Solidity code into a XML-based intermediate language that is later queried for specific syntactic patterns. Oyente~\cite{Luu16} uses symbolic execution to examine an under-approximation of some EVM bytecode; it can be fairly ineffective as far as code-coverage is concerned~\cite{Tsankov18}. EtherTrust~\cite{Grishchenko18a} and Securify~\cite{Tsankov18} are based on the generation of some symbolic constraints that soundly but coarsely over-approximate the behaviour of EVM bytecode; while EtherTrust uses horn clauses to represent this constraint, Securify uses a datalog program. They are intended to prove properties of programs rather than falsify them.

VerX~\cite{Permenev20} is a tool that uses symbolic execution and delayed predicate abstraction to check temporal (safety) properties of smart contracts. It is an automated tool that allows users to specify, using temporal logic~\cite{Manna95}, functional properties of smart contracts~\cite{Sergey18}; sometimes manual input is required to construct counterexamples. The solc-verify verifier~\cite{Hajdu19,Hajdu20} uses an approach similar to ours to create an \emph{unbounded} model checker for Solidity. It is based on a translation of Solidity into the Boogie language which is later checked by the Boogie verifier~\cite{Barnett05}. Given a Solidity program annotated with pre- and post-conditions, contract and loop invariants, and assertions, the verifier tries to discharge the proof obligations derived from the specification. This sort of modular/assume-guarantee reasoning can be quite precise and scalable but that, again, intrinsically depends on the ingenuity of the specifier. VeriSol~\cite{Wang18} is the framework that most resembles ours. It uses a translation from Solidity to Boogie that is later checked by Corral. It targets the contracts deployed to the Azure Blockchain and checks for \emph{semantic conformance} of contracts with respect to a workflow policy, which is expressed as a finite-state machine. VeriSol not only finds conformance violations but it derives invariants using predicate abstraction to automatically prove conformance.


\section{Formalising Solid and its execution environment}
\label{sec:Solid}

We formalise a representative subset of Solidity via the \emph{Solid} language. A Solid program is obtained by explicating/desugaring a Solidity program. It has the same constructs that Solidity has but slightly different semantics. For lack of space, we do not formalise the evaluation of expressions. We also formalise the behaviour of the \emph{Solid blockchain}: an execution environment that abstractly captures Ethereum's behaviour.

\subsection{Solidity and Solid}

\begin{figure}[!ht]
		\begin{grammar} \footnotesize
		<Program> := (contracts : <Contract>*)
		
		<Contract> := (types : <UserType>*, variables : <VariableDeclaration>*, \\ functions : <FunctionDefinition>*)
		
		<UserType> := {\bf Enum}(values : Name*) \alt {\bf Struct}(members : <VariableDeclaration>*)
		
		<VariableDeclaration> := (name : Name, type : <Type>)
		
		<FunctionDeclaration> := (visibility : VisibilitySpecifier, name : Name, \\ inParams : <LocalVariableDeclaration>*,\\ outParams : <LocalVariableDeclaration>*, body : <Statement>*) 
		
		<LocalVariableDeclaration> := (isPointer : bool, dataLocation : DataLocation, variableDeclaration : <VariableDeclaration>)
		
		<Type> ::=  {\bf UInt} | {\bf Int} | {\bf Bool} | {\bf Address} | {\bf ContractType}(name : Name) 
		\alt {\bf Array}(members : <Type>) | {\bf Mapping}(domain : <Type>, range : <Type>) \alt {\bf EnumType}(name : Name) | {\bf StructType}(name : Name) 
		
		<Statement> ::= {\bf WhileLoop}(condition : <Expression>, body : <Statement>*)
		\alt {\bf IfThenElse}(condition : <Expression>, ifBody : <Statement>*, \\ elseBody : <Statement>*) 
		\alt {\bf VariableDeclarationStatement}(variable : <LocalVariableDeclaration>) 
		\alt {\bf Assignment}(lhs : <Expression>, rhs : <Expression>) 
		\alt {\bf AllocMemory}(lhs : <Expression> , type : <Type>) | {\bf Revert} 
		\alt {\bf Require}(condition : <Expression>) | {\bf Return}(expression : <Expression>)
		\alt {\bf ContractCall}(lhs : <Expression>*, type : <Type>, funcName : Name, \\args : <Expression>*) 
		\alt {\bf CreateContract}(lhs : <Expression>, type : <Type>, args : <Expression>*) 
		\alt {\bf Transfer}(source : <Expression>, destination : <Expression>, \\value : <Expression>)
		\alt {\bf Send}(lhs : <Expression>, source : <Expression>, destination : <Expression>, \\value : <Expression>)
		\alt {\bf Call}(lhs : <Expression> , address : <Expression>, value : <Expression>)
		\alt {\bf VAssume}(condition : <Expression>) | {\bf VAssert}(condition : <Expression>)
	\end{grammar}
	\caption{Grammar of Solid}
	\label{fig:grammar}
\end{figure}

We introduce Solid (and Solidity) by presenting an abstract syntax for its constructs in Figure~\ref{fig:grammar}. We only introduce the Solidity/Solid constructs that are relevant for our formalisation; for a full account of Solidity see \cite{Solidity}. This grammar defines the abstract nodes of our syntax tree and their elements. Each production rule defines a node type and the constructors that can be used to create node instances of that type. For instance, the program rule defines that a program $p \in \textit{Program}$ is a $1$-tuple with a list of contracts given by its element $contracts$. We use the notation $n.el$ to access element $el$ of node $n$. For instance, $p.contracts$ gives the contracts of program $p$. Some nodes admit more than one constructor such as \textit{UserType}. For a node instance $n$ and constructor name $c$, the relation $n\isr{}c$ to establish whether $n$ was created using constructor $c$. Solid and Solidity share the same concrete syntax.

A smart contract is a stateful reactive program whose code and state are stored in the blockchain. It can execute its own internal code but it can also call/execute functions of other smart contracts. Smart contracts might also programmatically read addresses' balances and transfer currency. A \emph{Solidity/Solid program} is composed by a set of \emph{smart contracts}. A contract is similar to a class in object-oriented languages. It can declare a set of \emph{user-defined types}, a set of \emph{member variables} (we only consider private member variables), and a set of \emph{member functions}. While the member variables capture the (persistent) state of the contract, the functions describe its behaviour. All functions have an input parameter \tit{this} that designates the address on which this function is applied/called. For constructors, \tit{this} is instead a local variable denoting the address on which the new contract will be created. Functions have visibility modifiers to specify whether they are part of the public interface of the contract. For instance, the Solidity program in Fig.~\ref{fig:wallet} has a single contract \solidity{{Wallet}} that behaves like a toy bank: addresses can open and close an account, and deposit and withdraw money from their accounts.

\begin{figure}[ht]
	\begin{lstlisting}[numbers=none,basicstyle=\scriptsize\ttfamily]
contract Wallet {

	enum Status {
		None,
		Open,
		Closed
	}
	
	struct Account {
		uint id;
		uint balance;
		Status status;
	}
	
	mapping (address => Account) accounts;
	
	function open () public {
		require(accounts[msg.sender].status == Status.None);
		accounts[msg.sender].status = Status.Open;
	}
	
	function close () public {
		require(accounts[msg.sender].status == Status.Open);
		require(accounts[msg.sender].balance == 0);
		accounts[msg.sender].status = Status.Closed;
	}
	
	function deposit () payable public {
		require(accounts[msg.sender].status == Status.Open);
		accounts[msg.sender].balance = 
			accounts[msg.sender].balance + msg.value;
	}
	
	function withdraw (uint value) public {
		require(accounts[msg.sender].status == Status.Open);
		require(accounts[msg.sender].balance >= value);
		accounts[msg.sender].balance = 
			accounts[msg.sender].balance - value;
		msg.sender.transfer(value);
	}
}	
	\end{lstlisting}
	\caption{Wallet contract example.}
	\label{fig:wallet}
\end{figure}

Our formalisation supports the following basic types: address, contract, signed and unsigned 256-bit integers, and boolean. Both address and contract types are 160-bit unsigned integers that represent the identifier of an address in the blockchain. A contract type, however, is an identifier annotated/typed with the type of the contract that it stores. In addition to this basic types, we support arrays, mappings, and user-defined types. Recursive types, even mutually recursive, are not supported\footnote{In Solidity, a mapping breaks type recursion, i.e. if the recursion arises in the range of a mapping, then this is not considered a recursion. In Solid, any recursion is disallowed.}. Assignments and function calls cannot be used as expressions; programs that do not conform to that can be simply (and automatically) transformed into equivalent conformant ones.

A Solidity/Solid program has access to three (implicit) variables \solidity{{msg}}, \solidity{{tx}}, \solidity{{block}} that give dynamic information about function calls, transactions and blocks, respectively. Each function has as an input parameter a \solidity{{msg}} struct with two members: $\tit{sender}$ gives the address of the caller, and $\textit{value}$ the amount of currency transferred by the caller to the callee. A blockchain transaction has an associated \solidity{{tx}} structure with member $\textit{origin}$ which gives the address that issued the transaction. Finally, \solidity{{block}} is a structure with member $\tit{timestamp}$ giving the timestamp of the current block.

A Solidity/Solid program can inspect the balance of address $addr$ using the access expression $addr.balance$. Constructs \tbf{Transfer}, \tbf{Send} and \tbf{Call} can be used to transfer money between addresses; the latter also executes some function. \tbf{AllocMemory} allocates new value in memory. While \tbf{Revert} raises an exception, \tbf{Require} raises an exception if its condition is false. Exceptions cause the current call to be reverted and they are propagated upwards in the call stack; this propagation can be stopped by \tbf{Send} and \tbf{Call}.
%

The Solidity library \solidity{{Verification}} introduces verification primitives \solidity{{Assume(bool b)}} and \solidity{{Assert(bool b)}}. Calls to these functions are represented by the corresponding AST nodes \tbf{VAssume} and \tbf{VAssert}. We only support this specific Solidity library as a way to conveniently introduce our verification primitives. While \tbf{VAssume} ensures that its condition holds (by silently failing if it does not) at the point in the program in which it appear, \tbf{VAssert} raises an error.

\subsection{Explicating Solidity into Solid programs}

A Solid program is obtained by explicating/desugaring a Solidity program. It replaces some complex Solidity constructs by an equivalent set of simpler Solid statements. 

In Solidity, an assignment to the \tit{length} member of a dynamic array triggers a \emph{resizing}. When increasing the length of the array, only its length is altered. However, if the array's length decreases, the elements beyond the new length are reset to their \emph{default values}. Addresses, contracts, and integers have the integer literal zero as their default values, while \tit{false} is \tbf{Bool}'s default value. The default value for composite types is created by having its members set to their default values; fixed-sized arrays have their size constant while their data elements are set to their default values. Although mappings have a default value, they are not reset when such a resizing happens. 

Our explication replaces assignment \lsti{#array.length = #newSize}, where \lsti{#array} and \lsti{#newSize} are placeholders for an array expression and an integer expression, by the template:

\begin{lstlisting}[numbers=none,basicstyle=\scriptsize\ttfamily,mathescape=true]
if(#array.length > #newSize){
	uint i = #newSize;
	while(i < #array.length){
		$\tit{Reset}$(#array[i])
		i = i + 1;
	}
}
#array.length = #newSize;
\end{lstlisting}

\tit{Reset} is a meta-function that generates the code responsible for resetting element $i$ of the array. This code might involve a simple Solid assignment if the elements of the array are of a basic type; a list of assignments to reset its members if the element is a struct or a fixed-sized array; or even another block of statements of this kind if the elements of the array are themselves dynamic arrays. As types are not allowed to be recursive, this explication step is bound to terminate. The assignment to the array's length in the last statement of this block is a Solid assignment. Hence, it has no longer the original resizing semantics of Solidity but it merely changes the array's length; no element resetting occurs.

A contract has at its disposal a block of volatile storage spaces, called the \tit{memory}, which are allocated on demand as the contract needs more resources to execute; reference-type (i.e. structs and arrays) values are stored in it. Such values can also be stored in \tit{storage}: a persistent block of memory cells (disjoint from \tit{memory}) that stores the state of smart contracts. While an assignment between value-type expressions results in a simple copy, one between reference-type expressions can result in \emph{deep copying}, depending on where the values of these expressions are stored. Table~\ref{tab:deepcopy} depicts the semantics of assignments involving reference-type expressions. A reference-type value is represented by either a storage reference, or a storage or memory pointer. We assume the existence of a typing environment that captures this information for the expressions in each assignment. Roughly speaking, this environment marks local variables of a reference type declared as in $memory$ ($storage$) as storing a $memory$ ($storage$, respectively) pointer, whereas accessing a contract member variable (or their elements) is marked as a storage-reference expression. 

\begin{table}[ht]
	\caption{Solidity assignment semantics for reference types.}
		\label{tab:deepcopy}
	\centering
	\begin{tabular}{|c|c|c|}
		\hline
		Left-hand side & Right-hand side & Semantics \\
		\hline
		Storage reference & - & Deep copy \\ 
		\hline
		\multirow{3}{*}{Memory pointer} & Storage reference & Deep copy\\
														  \cline{2-3}
														  & Storage pointer & Deep copy\\
														  \cline{2-3}
														  & Memory pointer & Shallow copy (Aliasing)\\
		\hline
		\multirow{3}{*}{Storage pointer} & Storage reference & Shallow copy (Aliasing) \\
														\cline{2-3}
														& Storage pointer & Shallow copy (Aliasing) \\
														\cline{2-3}
														& Memory pointer & Deep copy \\
		\hline
	\end{tabular}\\
	The symbol - denotes that the type of the right-hand element is irrelevant.
\end{table}

Our explication process transforms deep-copy assignments into the element copies they correspond to. For structs and fixed-sized arrays, this step creates assignments between the corresponding elements of left-hand-side and right-hand-side expressions in the assignment. For dynamic arrays, a while loop carries the copy of data elements whereas a simple assignment resizes the left-hand-side array. Note that assignment between elements can again trigger deep copying and would need to be further explicated. For instance, let $\tit{lhs}$ and $\tit{rhs}$ be of type struct \lsti{SA} with members \lsti{uint a}, \lsti{uint[] b.} A deep copy between $\tit{lhs}$ and $\tit{rhs}$ would be explicated to \solidity{{lhs.a = rhs.a; lhs.b = rhs.b}}, then assignment \lsti{lhs.b = rhs.b} would be further explicated to:
\begin{lstlisting}[numbers=none,basicstyle=\scriptsize\ttfamily]
lhs.b.size = rhs.b.size;
uint i = 0;
while (i < lhs.b.size) {
	lhs.b[i] = rhs.b[i]
}
\end{lstlisting}

When the left-hand-side of the deep-copy assignment is in memory, an extra \tbf{AllocMemory} statement is added to allocate the space in memory for the new copy. For instance, if $\tit{lhs}$ was in memory, a \solidity{{lhs = new SA();}} would be added before the deep copying into \lsti{lhs} is carried out. A \solidity{{new}} statement is only used in Solidity to allocate dynamic arrays in memory but Solid extends its application to allocate any reference type in memory. Note here that the explication steps might depend on each other. For instance, \solidity{{lhs.b.size = rhs.b.size}} is an array-resizing assignment that needs to be explicated. Argument passing can also trigger deep copying. It can be refactored into assignment deep-copying by having arguments assigned to a temporary variable of the same type as the parameter they correspond to. One can think of all of these steps as being applied to the program until no further explication is possible.

Aside from eliminating deep copies and array resizings, our explication process adds preconditions to expressions that might have an undefined behaviour/trigger some exception. For instance, an array access beyond the array's length throws an exception and, so, we precede array access  \solidity{{#array[i]}} by \solidity{{require(i < #array.length)}}. It also explicates array-\emph{push} operations, adds precondition \solidity{{require(msg.value == 0)}} to non-payable functions, and initial assignments/allocations to functions to set local variables and return parameters to their default values.

\subsection{The Solid blockchain}

The Solid blockchain has a collection of addresses, for which states are captured by \emph{address mapping} $\ttbf{s}$. \emph{Address cell} $\ttbf{s}[addr]$ stores the state of address $addr \in \tit{Address}$. As a convention we use the italic name of a type to represent the set of values it is associated with. For instance, while \tbf{Address} gives a type constructor in our language, \tit{Address} gives the set of unsigned 160-bit integers. An address cell is a record with 4-elements: $\tit{type} \in \tit{AddressType}$ gives the address' type, $\tit{balance} \in \tit{UInt}$ gives its balance, $\tit{members}$ and $\tit{storage}$ are only relevant when the address represents a smart-contract instance; they store the state of the instance located at the corresponding address. An address' type can be \tit{Unused}, \tit{SimpleAddress}, or a particular contract's name in \tit{Contracts}. Value \textit{Unused} represents unused/uncontrolled addresses; these addresses neither have a secret key or contract code to control them. \textit{SimpleAddress} represents a simple address whose balance is controlled by a secret key, whereas a name in \tit{Contracts} means that the corresponding address is an instance of the given contract (and controlled by its code). We capture the state of Solid's \solidity{{block}} variable by $\ttbf{b}$: a record with a single element $\tit{time} \in \tit{UInt}$ where $\tit{time}$ gives the timestamp of the current block. Solid blockchain's state is given by $(\ttbf{s},\ttbf{b})$.

The element $\tit{storage}$ of an address cell is a \emph{reference mapping}: $\tit{storage}[r]$ gives the \tit{reference cell} that $r$ points to. A reference cell is a record with two elements: $\tit{type} \in \tit{Type} \cup \{\tit{None}\}$ the Solid type of the value stored, and $\tit{value}$ the actual value stored; $\tit{value}$ in the disjoint union of $\{\tit{UInt}, \tit{Int}, \tit{Bool}, \tit{Address}, \tit{Ref}\} \cup \{t \in \tit{Enumerations} \cup \tit{Records} \cup \tit{Mappings}\}$. \tit{Enumerations}, \tit{Records}, \tit{Mappings} capture enumerations (by the sequence of its member values), records (as tuples with named members), and total mappings. The reference type $\tit{Ref}$ holds an infinite set of special values that can be tested for equality. \tit{None} is a semantic element such that $\tit{type} = \tit{None}$ captures that the reference has not been allocated yet. The identifier mapping $\tit{members}$ captures where member variable values are stored: $\tit{members}[m]$ gives the reference where $m$ is stored in $\tit{storage}$.

The state of a Solid variable of type $T$ is represented in a reference mapping by a cell with $\tit{value}$ in type $\tit{RefType}(T)$. The state of a variable of type \tbf{UInt}, \tbf{Int}, or \tbf{Bool} is captured using \tit{UInt}, \tit{Int}, or \tit{Bool}, respectively. The state of a variable of type \tbf{Address} or \tbf{ContractType} is represented by \tit{Address}. \tbf{Enum} definitions in a Solid program give rise to a similar enumeration definition in \tit{Enumerations}, and their state is represented by a value of this corresponding enumeration. $\tit{RefType}(\tbf{Array}(T))$ is a record with members $\tit{length} \in \tit{UInt}$ and $\tit{data} \in \tit{UInt} \rightarrow T'$, where $\tit{UInt} \rightarrow T'$ is a total mapping from $\tit{UInt}$ to $T'$, and if $T$ is a reference type (i.e. $T$ is a Solid array or struct) then $T'$ is \tit{Ref}, otherwise $T'$ is $\tit{RefType}(T)$. In general, reference-type elements of a composite type are represented by references that point to reference cells storing the members' state, whereas non-reference-type elements are simply represented by their value. So, a struct is represented by a record where each member $m$ of type $T$ gives rise to a member $m$ of type $\tit{Ref}$ if $T$ is a reference type, and $\tit{RefType}(T)$ otherwise. $\tbf{Mapping}(D,R)$ is represented  by a total mapping $\tit{RefType}(D) \rightarrow R'$ where $R'$ is \tit{Ref} or $\tit{RefType}(R)$ according to whether or not $R$ is a reference type. The domain type of this mapping is given by $\tit{RefType}(D)$ as we only allow Solid mappings to have an elementary domain type. For instance, the state of a variable of type \solidity{{S}} where \solidity{|struct S {uint a; uint[] b;}|} is captured by a value of record type $\tit{RefType}(S)$ with members $a \in \tit{UInt}$ and $b \in \tit{Ref}$ where $b$ points to a record value representing $\tbf{Array}(\tbf{UInt})$. The state of a local variable or parameter is captured in a local context by a value of type $\tit{LocalType}(T)$. For a reference type $T$, $\tit{LocalType}(T) = \tit{Ref}$ and the reference value points to a value in $\tit{RefType}(T)$, whereas, for a value type $T$, $\tit{LocalType}(T) = \tit{RefType}(T)$.

The Solid blockchain processes the following transactions:

\begin{itemize}
	\item $\textbf{create-address}(\tit{value})$ creates an address with $\tit{value} \in \tit{UInt}$ as balance;
	\item $\textbf{currency-transfer}(\tit{src}, \tit{dest}, \tit{value})$ transfers $\tit{value} \in \tit{UInt}$ from $\tit{src} \in \tit{Address}$ to $\tit{dest} \in \tit{Address}$.
	\item $\textbf{create-contract}(\tit{src}, \tit{type}, \tit{value}, \tit{args})$ creates an instance of contract $\tit{type} \in \textit{Contracts}$ with balance $\tit{value} \in \tit{UInt}$ and call contract $\tit{type}$'s constructor with arguments $\tit{args}$ - this transaction is issued by address $\tit{src}$;
	\item $\textbf{execute-contract}(\tit{src}, \tit{addr}, \tit{type}, \tit{func}, \tit{value}, \tit{args})$ executes function $\tit{func}$ of contract $\tit{type} \in \textit{Contracts}$ deployed at address $\tit{addr}$ with arguments $\tit{args}$ - this transaction is issued by address $\tit{src}$;
	\item $\textbf{mint-block}$ is a control function that denotes that a block has been mined/completed.
\end{itemize}

\begin{table}[ht]
	\setlength{\tabcolsep}{0pt}
	\centering
	\caption{Operational semantics of transactions.}
	\label{tab:transactions}
	\resizebox{.7\textwidth}{!}{%
	\begin{tabular}{| c | c |}
		\hline 
		Premise & Conclusion \\
		\hline
		$\begin{array}{c} tr \isr \textbf{create-address}, \\ \ttbf{s}[\tit{addr}^{\bullet}].\tit{type} = \textit{Unused} \end{array}$ & $\begin{array}{l}  \ttbf{s}'[\tit{addr}^{\bullet}].\tit{type} = \\ \quad \textit{SimpleAddress}, \\ \ttbf{s}'[\tit{addr}^{\bullet}].balance = \\ \quad tr.value \end{array}$ \\
		\hline
		$\begin{array}{c} tr \isr \textbf{currency-transfer}, \\ \ttbf{s}[tr.src].\tit{type} = \textit{SimpleAddress},\\ \ttbf{s}[tr.de\tit{st}].\tit{type} \in \{\textit{SimpleAddress}\} \cup \textit{Contracts},\\ \ttbf{s}[tr.src].balance \geq tr.value \end{array}$ &  $ \begin{array}{l} \ttbf{s}' = \textit{set_bal}(\ttbf{s}, tr.src, \\ \quad tr.de\tit{st}, tr.value) \end{array}$ \\
		\hline
		$\begin{array}{c} tr \isr \textbf{create-contract},\\ \ttbf{s}[\tit{addr}^{\bullet}].\tit{type} = \textit{Unused}, \\ \ttbf{s}^{\circ} = \textit{set_bal}(\ttbf{s}, tr.src, \tit{addr}^{\bullet}, tr.value), \\ \ttbf{s}^{\bullet} = \textit{init_s}(\ttbf{s}^{\circ},\tit{addr}^{\bullet}, \tit{st}.\tit{type}), \\ \ttbf{tx}^{\bullet}.origin = tr.src, \\ f^{\bullet} = \textit{constructor}(\tit{st}.\tit{type}), \\ \ttbf{m}^{\bullet} = \tit{zero_m}, \\ \ttbf{l}^{\bullet} = \textit{init_l}(f^{\bullet}, tr.args), \\ \ttbf{c}^{\bullet} = f^{\bullet}.body, \\ (\ttbf{s}^{\bullet}, \ttbf{b}, \ttbf{tx}^{\bullet}, \ttbf{m}^{\bullet}, \ttbf{l}^{\bullet}, \ttbf{c}^{\bullet}) \spath{}^{*} \qquad\qquad \\ \qquad\qquad\qquad  (\ttbf{s}^{\diamond}, \ttbf{b}, \ttbf{tx}^{\bullet}, \ttbf{m}^{\diamond}, \ttbf{l}^{\diamond}, \emptySeq) \end{array}$ & $\ttbf{s}' = \ttbf{s}^{\diamond}$  \\
		\hline
		$\begin{array}{c} tr \isr \textbf{execute-contract},\\ \ttbf{s}[tr.address].\tit{type} = tr.\tit{type},  \\ \ttbf{s}^{\bullet} = \textit{set_bal}(\ttbf{s}, tr.src, tr.de\tit{st}, tr.value), \\\ttbf{tx}^{\bullet}.origin = tr.src, \\  f^{\bullet} = \textit{function}(tr.\tit{type}, \textit{tr.functionName}), \\ \ttbf{m}^{\bullet} = \tit{zero_m}, \\ \ttbf{l}^{\bullet} = \textit{init_l}(f^{\bullet}, tr.args), \\ \ttbf{c}^{\bullet} = f^{\bullet}.body, \\ (\ttbf{s}^{\bullet}, \ttbf{b}, \ttbf{tx}^{\bullet}, \ttbf{m}^{\bullet}, \ttbf{l}^{\bullet}, \ttbf{c}^{\bullet}) \spath{}^{*} \qquad\qquad \\ \qquad\qquad\qquad  (\ttbf{s}^{\diamond}, \ttbf{b}, \ttbf{tx}^{\bullet}, \ttbf{m}^{\diamond}, \ttbf{l}^{\diamond}, \emptySeq) \end{array}$ &  $\ttbf{s}' = \ttbf{s}^{\diamond}$\\
		\hline
		$\begin{array}{l} tr \isr \textbf{mint-block}, \\ time^{\bullet} > \ttbf{b}.time \end{array}$  & $\ttbf{b}'.time = time^{\bullet} $  \\
		\hline
	\end{tabular}}
\end{table} 

Table~\ref{tab:transactions} presents the SOS (Structural operational semantics) rules~\cite{Plotkin81} that formalise how each of these transactions alter the blockchain's state. The variables superscripted with $\bullet$, $\circ$, and $\diamond$ are implicitly existentially quantified; the quantification is left out to make our presentation less cluttered. Each rule defines a premise and a conclusion. If a transaction $tr$ and state $(\ttbf{s},\ttbf{b})$ satisfy the premise, the blockchain can transition to a state $(\ttbf{s}', \ttbf{b}')$ that satisfies the conclusion; such a transition is represented by $(\ttbf{s},\ttbf{b}) \Path{t} (\ttbf{s}',\ttbf{b}')$. The elements of post-state that are not constrained in the conclusion are left unchanged with respect to the pre-state. We denote the path $(\ttbf{s},\ttbf{b}) \Path{t_1} (\ttbf{s}_1,\ttbf{b}_1) \Path{t_2} \ldots \Path{t_{n-1}}  (\ttbf{s}_{n-1},\ttbf{b}_{n-1}) \Path{t_n} (\ttbf{s}',\ttbf{b}')$ by $(\ttbf{s},\ttbf{b}) \Path{t_1,\ldots,t_n} (\ttbf{s}',\ttbf{b}')$. 

For address mapping $\ttbf{s}$, the function $\tit{set_bal}(\ttbf{s}, \tit{src}, \tit{dest}, \tit{val})$ gives the address mapping $\ttbf{s}'$ resulting from transferring $\tit{val}$ from the balance of $\tit{src}$ to the balance of $\tit{dest}$ in $\ttbf{s}$. Function $\tit{constructor}(\tit{type})$ gives the constructor of contract $\tit{type}$, whereas $\tit{function}(\tit{type}, \tit{name})$ gives the function $\tit{name}$ of contract $\tit{type}$. The value \tit{zero_m} gives reference map where all references are uninitialised, i.e. they point to \tit{None}. The function $\tit{init_s}(\ttbf{s}, \tit{addr}, \tit{type})$ modifies $\ttbf{s}$ to store and initialise an instance of smart contract $\tit{type}$ in address $\tit{addr}$. For address cell at $\tit{addr}$, its type is set to $\tit{type}$, balance is set to 0, and its member variables are allocated and initialised - each member variable $m$ of type $T$ gives rise to $m \mapsto r$ in $\tit{members}$ and reference $r$ is allocated and initialised in the address' storage to capture the default value of $T$. This process allocates a tree of reference cells to capture the state of this member as per $\tit{RefType}(T)$. The allocation procedure precludes two distinct reference-type members/elements from pointing to the same reference cell. We further discuss this allocation process in Sec.~\ref{sec:bmc}. The function $\tit{init_l}(f, args)$ creates a local context (an identity mapping) with all variables declared in $f$ with input parameters initialised to their corresponding \tit{args} value whereas the other variables are non-deterministically initialised; note that our explication procedure creates initial statements in the function $f$ to properly initialise and allocate these variable to their default value. The processing of \tbf{create-contract} and \tbf{execute-contract} transactions involve the execution of a smart-contract function, captured by predicate $x \spath{}^{*} x'$.

These rules make some abstractions on the actual behaviour of Ethereum/Solidity. Ethereum addresses cannot be created with a predetermined balance as per \tit{create-address} transaction. This abstraction captures the fact that addresses and contracts can receive currency via, for instance, a coinbase transaction~\cite{Solidity}. 
Furthermore, our formalisation makes a fundamentally different design choice in modelling the execution of smart contracts: it uses contract types to validate executions; as per, for instance, \tbf{execute-contract} rule. Solidity executes a function if its signature matches the signature of the function being called, even if the type of the contract being called is completely different from what the caller expects. Hence, a function with the same signature but with a completely different behaviour can be executed. Even worse than that, if no function matches a given signature, the fallback function deployed in the address called is executed. To avoid this sort of behaviour, the Solid blockchain relies on the \tit{type} element of an address cell to specify and check the type of a deployed contract, blocking calls to contract instances of a wrong type.

A smart-contract \tit{execution state} is a 5-tuple $(\ttbf{s}, \ttbf{b}, \ttbf{tx}, \ttbf{m}, \ttbf{l}, \ttbf{c})$ where $\ttbf{s}$ is an address mapping, $\ttbf{b}$ is a block-information structure, $\ttbf{tx}$ is a transaction-information record with one element $\tit{origin}$ (capturing the state of Solid's variable \solidity{{tx}}) that gives the address that has issued the transaction, $\ttbf{m}$ is a reference mapping that captures the state of the memory, $\ttbf{l}$ is a local identifier mapping capturing the value of local variables and parameters, and $\ttbf{c}$ is a piece of code to be executed. The execution of a smart contract alters the execution state according to the SOS rules depicted in Tables \ref{tab:simple-statements} and \ref{tab:function-call-statements}. If execution state $x = (\ttbf{s}, \ttbf{b}, \ttbf{tx}, \ttbf{m}, \ttbf{l}, \ttbf{c})$ satisfies the premise of a rule, a transition to a state $x' = (\ttbf{s}', \ttbf{b}', \ttbf{tx}', \ttbf{m}', \ttbf{l}', \ttbf{c}')$ satisfying the conclusion can be created; such a transition is denoted by $x \spath{} x'$. Our rules assume that $\ttbf{c} = \Seq{st}\cat \tit{tail}$, variables superscripted with $\bullet$, $\circ$, and $\diamond$ are implicitly existentially quantified, and the elements of the post-state that are not constrained in the conclusion are left unchanged with respect to the pre-state. Moreover, we use $x \spath{}^{*} x'$ to denote the existence of a path of transitions (i.e. an execution) from execution state $x$ to $x'$. The transition relation $\spath{}^{*}$ is reflexive. 

\begin{table}
	\centering
	\caption{Operational semantics of simple statements.}
	\label{tab:simple-statements}
	\setlength{\tabcolsep}{1pt}
	\resizebox{.75\textwidth}{!}{%
	\begin{tabular}{| c | l |}
		\hline
		 Premise & \hspace{2pt} Conclusion \\
		\hline
		 $\tit{st} \isr{} \tbf{WhileLoop}$, $\tit{st}.\tit{condition}$ & \hspace{2pt} $\ttbf{c}' = \tit{st}.\tit{body}  \cat \ttbf{c}$  \\
		\hline
		 $\tit{st} \isr{} \tbf{WhileLoop}$, $\lnot \tit{st}.\tit{condition}$ & \hspace{2pt} $\ttbf{c}' = \tit{tail}$  \\
		\hline
		 $\tit{st} \isr{} \tbf{IfThenElse}$, $ \tit{st}.\tit{condition}$ & \hspace{2pt}  $\ttbf{c}' = \tit{st}.\tit{ifBranch}  \cat \tit{tail}$  \\
		\hline
		 $\tit{st} \isr{} \tbf{IfThenElse}$, $\lnot \tit{st}.\tit{condition}$ & \hspace{2pt} $\ttbf{c}' = \tit{st}.\tit{elseBranch}  \cat \tit{tail}$  \\
	    \hline
		 $\tit{st} \isr{} \tbf{VariableDeclarationStatement}$  & \hspace{2pt}  $\ttbf{c}' = \tit{tail}$  \\		
		\hline
	   	 $\tit{st} \isr{} \tbf{Assignment}$ &  \hspace{2pt} $\arraycolsep=0pt\begin{array}{l}(\ttbf{s}', \ttbf{m}', \ttbf{l}') =\begin{array}[t]{l}\tit{update}(\Seq{\tit{st}.\tit{lhs}},\Seq{\tit{st}.\tit{rhs}}, \\ 
	   	 					\quad \ttbf{s},\ttbf{b},\ttbf{tx}, \ttbf{m}, \ttbf{l},\ttbf{l}), \end{array} \\ \ttbf{c}' = \tit{tail} \end{array}$  \\
	   	\hline
   		$\tit{st} \isr{} \tbf{Return}$  &  \hspace{2pt} $\arraycolsep=0pt\begin{array}{l} (\ttbf{s}', \ttbf{m}', \ttbf{l}') =\begin{array}[t]{l}\tit{update}(\tit{outs}(f),\tit{st}.\tit{exprs}, \\ \quad \ttbf{s},\ttbf{b}, \ttbf{tx}, \ttbf{m}, \ttbf{l},\ttbf{l}), \end{array}\\ \ttbf{c}' = \emptySeq \end{array}$  \\
   		\hline
   		 $\begin{array}{c} \tit{st} \isr{} \tbf{AllocMemory}, \\ unallocated(\ttbf{m},\tit{ref}^{\bullet},\tit{st}.\tit{type}) \end{array}$ & \hspace{2pt} $\arraycolsep=0pt\begin{array}{l} m^{\bullet} = alloc(\ttbf{m},\tit{ref}, \tit{st}.\tit{type}), \\ (\ttbf{s}', \ttbf{m}', \ttbf{l}') =\begin{array}[t]{l}\tit{update}(\Seq{\tit{lhs}}, \Seq{\tit{ref}^{\bullet}}, \ttbf{s},\ttbf{b}, \\ \quad \ttbf{tx},m^{\bullet}, \ttbf{l}, \ttbf{l}),~ \end{array}\\ \ttbf{c}' = \tit{tail} \end{array}$\\
   		\hline
   		 $\begin{array}{c} \tit{st} \isr{} \tbf{Revert} \end{array}$   & \hspace{2pt}  $\ttbf{c}' = \Seq{\ttbf{fail}}$  \\
   		\hline
   		 $\begin{array}{c} \tit{st} \isr{} \tbf{Require}, \tit{st}.\tit{condition} \end{array}$   & \hspace{2pt} $\ttbf{c}' = \tit{tail}$  \\
   		\hline
   		 $\begin{array}{c} \tit{st} \isr{} \tbf{Require}, \lnot \tit{st}.\tit{condition} \end{array}$   & \hspace{2pt}  $\ttbf{c}' = \Seq{\ttbf{fail}}$  \\
   		\hline
   		 $\begin{array}{c} \tit{st} \isr{} \tbf{Transfer}, \\ \ttbf{s}[\tit{st}.\tit{destination}].\tit{type} = \textit{Address} \end{array}$  & \hspace{2pt}  $\arraycolsep=0pt\begin{array}{l} \ttbf{s}' =\begin{array}[t]{l}\textit{set_bal}(\ttbf{s}, \tit{st}.source, \\ \quad  \tit{st}.\tit{destination}, \tit{st}.value), \end{array}\\ \ttbf{c}' = \tit{tail} \end{array}$\\
   		\hline
   		 $\begin{array}{c} \\ \tit{st} \isr{} \tbf{Transfer}, \\ \ttbf{s}[\tit{st}.\tit{destination}].\tit{type} \in \textit{Contracts} \end{array}$  & \hspace{2pt}  $\arraycolsep=0pt\begin{array}{l} \ttbf{s}' =\begin{array}[t]{l}\textit{set_bal}(\ttbf{s}, \tit{st}.source, \\ \quad \tit{st}.\tit{destination}, \tit{st}.value), \end{array} \\ \ttbf{c}' = \tit{tail} \end{array}$\\
   		\cline{2-2}
   		 $\begin{array}{c} ~ \\ ~ \end{array}$ &  \hspace{2pt} $\ttbf{c}' = \Seq{\ttbf{fail}}$  \\
   		\hline
   		 $\begin{array}{c} \tit{st} \isr{} \tbf{Send}, \\ \ttbf{s}[\tit{destination}].\tit{type} = \textit{Address} \end{array}$  & \hspace{2pt}  $\arraycolsep=0pt\begin{array}{l} \ttbf{s}^{\bullet} =\begin{array}[t]{l}\textit{set_bal}(\ttbf{s}, \tit{st}.source, \\ \quad \tit{st}.\tit{destination}, \tit{st}.value), \end{array} \\ (\ttbf{s}', \ttbf{m}', \ttbf{l}') =\begin{array}[t]{l}\tit{update}(\tit{lhs},\Seq{\ttbf{True}}, \\ \quad\ttbf{s}^{\bullet},\ttbf{b},\ttbf{tx}, \ttbf{m}, \ttbf{l},\ttbf{l}), \end{array} \\ \ttbf{c}' = \tit{tail} \end{array}$\\
   		\hline
   		 $\begin{array}{c} \\ \tit{st} \isr{} \tbf{Send}, \\ \ttbf{s}[\tit{destination}].\tit{type} \in \textit{Contracts} \end{array}$  & \hspace{2pt}  $\arraycolsep=0pt\begin{array}{l}  \ttbf{s}^{\bullet} = \begin{array}[t]{l} \textit{set_bal}(\ttbf{s}, \tit{st}.source, \\ \quad \tit{st}.\tit{destination}, \tit{st}.value), \end{array} \\ (\ttbf{s}', \ttbf{m}', \ttbf{l}') = \begin{array}[t]{l} \tit{update}(\tit{lhs},\Seq{\ttbf{True}}, \\ \quad \ttbf{s}^{\bullet},\ttbf{b},\ttbf{tx}, \ttbf{m}, \ttbf{l},\ttbf{l}), \end{array} \\ \ttbf{c}' = \tit{tail} \end{array}$\\
   		\cline{2-2}
   		  $\begin{array}{c} ~ \\ ~ \end{array}$ & \hspace{2pt}  $\arraycolsep=0pt\begin{array}{l} (\ttbf{s}', \ttbf{m}', \ttbf{l}') = \begin{array}[t]{l} \tit{update}(\tit{lhs},\Seq{\ttbf{False}}, \\ \quad \ttbf{s},\ttbf{b},\ttbf{tx}, \ttbf{m}, \ttbf{l},\ttbf{l}), \end{array} \\ \ttbf{c}' = \tit{tail} \end{array}$  \\
   		\hline
   		$\begin{array}{c} \tit{st} \isr{} \tbf{VAssume} \end{array}$   & \hspace{2pt} $\ttbf{c}' = \Seq{\ttbf{fail}}$  \\
   		\hline
   		$\begin{array}{c} \tit{st} \isr{} \tbf{VAssert} \end{array}$   & \hspace{2pt}  $\ttbf{c}' = \Seq{\ttbf{error}}$  \\
   		\hline
	\end{tabular}}
\end{table} 

The rules for simple statements in Table \ref{tab:simple-statements} propose loop unfolding to capture the behaviour of a while-loop and branch selection for an if-then-else. We use expressions in some premises to denote their evaluation in the current state. Some of our rules rely on function $\tit{update}(\Seq{\tit{lhs}_1,\ldots, \tit{lhs}_n}, \Seq{\tit{rhs}_1,\ldots, \tit{rhs}_n}, \ttbf{s}, \ttbf{b}, \ttbf{tx}, \ttbf{m}, \ttbf{l}, \ttbf{l})$. It updates the value of expressions $\tit{lhs}_i$ to the value of $\tit{rhs}_i$. While $\tit{lhs}_i$ is evaluated with respect to $\ttbf{s}$, $\ttbf{b}$, $\ttbf{tx}$, $\ttbf{m}$, and $\ttbf{l}$, $\tit{rhs}_i$ is evaluated with respect to $\ttbf{s}$, $\ttbf{b}$, $\ttbf{tx}$, $\ttbf{m}$, and $\ttbf{l}'$. This evaluation of expressions with different local contexts is relevant for capturing the return values of a function call. Depending on whether $\tit{lhs}_i$ represents a location in storage, memory or in the local context, this function updates $\ttbf{s}$, $\ttbf{m}$, or $\ttbf{l}$, respectively. In the \tbf{Return} rule, $\tit{outs}(f)$ gives the list of output-parameter identifiers for the function $\tit{f}$ containing the return command.

The \tbf{AllocMemory} statement relies on predicates $\tit{unallocated}$ and $\tit{alloc}$ to create of new values in the memory. While $\tit{unallocated}(\ttbf{m}, \tit{ref}, \tit{tp})$ ensures that, in reference mapping $\ttbf{m}$, reference \tit{ref} points to a tree of unallocated reference cells (i.e. every cell's $\tit{type}$ is \tit{None}) capturing the structure of a value of type $\tit{tp}$. As for \tit{init_s}, this predicate also stipulates that two distinct reference elements in this tree point to different reference cells.
 The $\tit{alloc}(\ttbf{m}, \tit{ref}, \tit{tp})$ function modifies $\ttbf{m}$ so that all elements in the tree of reference cells pointed to by $\tit{ref}$ have their appropriate $\tit{type}$, and $\tit{value}$ is initialised to their default value, representing an initial value of type $\tit{tp}$; reference values are left unchanged. 


The semantic commands \ttbf{fail} and \ttbf{error} cannot appear in Solid code and serve to indicate that a piece of code silently terminated due to some failure and that some error has been raised, respectively. The rules for constructs \tbf{Transfer} and \tbf{Send} when $\tit{destination}$ is a contract have two different conclusions. They represent two distinct rules with the same premise but different conclusions. The execution of these statements might fail or return a different value due to conditions that we have abstracted away in our formalisation. Hence, the non-deterministic behaviour captured by these rules.

\begin{table}[t]
	\centering
	\setlength{\tabcolsep}{2pt}
	\caption{Operational semantics of function-call statements.}
	\label{tab:function-call-statements}
\centerline{\resizebox{1.1\textwidth}{!}{%
\begin{tabular}{| c | c |}
		\hline
		Premise & Conclusion \\
		\hline
		$\begin{array}{c} \tit{st} \isr{} \tbf{ContractCall}, \\ \ttbf{s}[\tit{st.args.this}].\tit{type} = \tit{st.type},  \\ \ttbf{s}^{\bullet} = \textit{set_bal}(\ttbf{s}, \tit{this}, \tit{st.args.this}, \tit{st.args.msg.value}), \\ f^{\bullet} = \textit{function}(\tit{st.type}, \textit{st.functionName}) \\ \ttbf{l}^{\bullet} = \textit{init_l}(f^{\bullet}, \tit{st.args}), \\ \ttbf{c}^{\bullet} = f^{\bullet}.\tit{body}, \\ (\ttbf{s}^{\bullet}, \ttbf{b}, \ttbf{tx}, \ttbf{m}, \ttbf{l}^{\bullet}, \ttbf{c}^{\bullet}) \spath{}^{*} (\ttbf{s}^{\diamond}, \ttbf{b}, \ttbf{tx}, \ttbf{m}^{\diamond}, \ttbf{l}^{\diamond}, \ttbf{c}^{\diamond}) \end{array}$   &  $\begin{array}{l} 
			\text{For }\ttbf{c}^{\diamond} = \emptySeq\text{:} \\ 
				\qquad (\ttbf{s}', \ttbf{m}', \ttbf{l}') = \tit{update}(\tit{lhs},\tit{outs}(f^{\bullet}),\ttbf{s}^{\diamond},\ttbf{b},\ttbf{tx}, \ttbf{m}^{\diamond}, \ttbf{l},\ttbf{l}^{\diamond}), \\ 
				\qquad \ttbf{c}' = \tit{tail}  \\
			\text{For }\ttbf{c}^{\diamond} \in \{ \Seq{\ttbf{fail}},\Seq{\ttbf{error}}\}\text{:} \\ 
				\qquad \ttbf{c}' = \ttbf{c}^{\diamond}
			\end{array}$  \\
		\hline
		$\begin{array}{c} \tit{st} \isr{} \tbf{Call}, \\ \ttbf{s}[\tit{st.args.this}].\tit{type} = \textit{SimpleAddress} \end{array}$   &  $\begin{array}{c}  \ttbf{s}^{\bullet} = \textit{set_bal}(\ttbf{s}, \tit{this}, \tit{st.args.this}, \tit{st.args.msg.value}), \\ (\ttbf{s}', \ttbf{m}', \ttbf{l}') = \tit{update}(\tit{lhs},\Seq{\ttbf{True}},\ttbf{s}^{\bullet},\ttbf{b},\ttbf{tx}, \ttbf{m}, \ttbf{l},\ttbf{l}), \\ \ttbf{c}' = \tit{tail} \end{array}$\\
		\hline
		$\begin{array}{c} st \isr{} \tbf{Call}, \\ \ttbf{s}[\tit{st.args.this}].\tit{type} \in \textit{Contracts}, \\ \ttbf{s}^{\bullet} = \textit{set_bal}(\ttbf{s}, \tit{this}, \tit{st.address}, \tit{st.args.msg.value}), \\f^{\bullet} \in \textit{functions}(\ttbf{s}[\tit{st.args.this}].\tit{type}), \\ \ttbf{l}^{\bullet} = \textit{init_l}(f^{\bullet}, \tit{st.args}), \\ \ttbf{c}^{\bullet} = f^{\bullet}.\tit{body}, \\ (\ttbf{s}^{\bullet}, \ttbf{b}, \ttbf{tx}, \ttbf{m}, \ttbf{l}^{\bullet}, \ttbf{c}^{\bullet}) \spath{}^{*} (\ttbf{s}^{\diamond}, \ttbf{b},\ttbf{tx}, \ttbf{m}^{\diamond}, \ttbf{l}^{\diamond}, \ttbf{c}^{\diamond}) \end{array}$   &  $\begin{array}{l} 
			\text{For }\ttbf{c}^{\diamond} = \emptySeq\text{:} \\ 
				\qquad (\ttbf{s}', \ttbf{m}', \ttbf{l}') = \tit{update}(\tit{lhs},\Seq{\ttbf{True}},\ttbf{s}^{\diamond},\ttbf{b},\ttbf{tx}, \ttbf{m}^{\diamond}, \ttbf{l},\ttbf{l}^{\diamond}),\\ 
				\qquad \ttbf{c}' = \tit{tail} \\
			\text{For }\ttbf{c}^{\diamond} = \Seq{\ttbf{fail}}\text{:} \\
				\qquad (\ttbf{s}', \ttbf{m}', \ttbf{l}') = \tit{update}(\tit{lhs},\Seq{\ttbf{False}},\ttbf{s},\ttbf{b},\ttbf{tx}, \ttbf{m}, \ttbf{l},\ttbf{l}), \\ 
				\qquad \ttbf{c}' = \tit{tail} \\
			\text{For }\ttbf{c}^{\diamond} = \Seq{\ttbf{error}}: \\
				\qquad \ttbf{c}' = \Seq{\ttbf{error}}
		\end{array}$\\
		\hline
		$\begin{array}{c} \tit{st} \isr{} \tbf{CreateContract},\\ \ttbf{s}[\tit{addr}^{\bullet}].\tit{type} = \textit{None}, \\ \ttbf{s}^{\circ} = \textit{set_bal}(\ttbf{s}, \tit{this}, \tit{addr}^{\bullet}, \tit{st.args.msg.value}), \\ \ttbf{s}^{\bullet} = \textit{init_s}(\ttbf{s}^{\circ},\tit{addr}^{\bullet}, \tit{st.type}), \\ f^{\bullet} = \textit{constructor}(\tit{st.type}), \\ \ttbf{l}^{\bullet} = \textit{init_l}(f^{\bullet}, \tit{st.args}), \\ \ttbf{c}^{\bullet} = f^{\bullet}.\tit{body}, \\ (\ttbf{s}^{\bullet}, \ttbf{b}, \ttbf{tx}, \ttbf{m}, \ttbf{l}^{\bullet}, \ttbf{c}^{\bullet}) \spath{}^{*} (\ttbf{s}^{\diamond}, \ttbf{b}, \ttbf{tx}, \ttbf{m}^{\diamond}, \ttbf{l}^{\diamond}, \ttbf{c}^{\diamond}) \end{array}$ 
		&  $\begin{array}{l} 
			\text{For }\ttbf{c}^{\diamond} = \emptySeq\text{:} \\ 
				\qquad (\ttbf{s}', \ttbf{m}', \ttbf{l}') = \tit{update}(\Seq{\tit{lhs}},\Seq{addr^{\bullet}}, \ttbf{s}^{\diamond},\ttbf{b}, \ttbf{tx}, \ttbf{m}^{\diamond}, \ttbf{l},\ttbf{l}^{\diamond}),\\ 
				\qquad \ttbf{c}' = \tit{tail} \\
			\text{For }\ttbf{c}^{\diamond} \in \{ \Seq{\ttbf{fail}},\Seq{\ttbf{error}}\}\text{:} \\ 
				\qquad \ttbf{c}' = \ttbf{c}^{\diamond}
			\end{array}$  \\
		\hline
	\end{tabular}}}
\end{table}

The rules in Table \ref{tab:function-call-statements} deal with function calls. Some rules have different conclusions according to a condition on $\ttbf{c}^{\diamond}$; this notation concisely represents different rules with similar premises. All these statements have different behaviours according to whether the underlying function called executes successfully or fails. The \tbf{ContractCall} statement conforms to our design choice of checking contract types when a smart-contract function is called. The \tbf{Call} primitive can be used to bypass this check. In Solidity, this primitive triggers a specific function according to a function signature passed as a parameter. In our formalisation, however, we ignore this function-signature parameter and assume that \tbf{Call} chooses non-deterministically a function $\tit{f}^{\bullet}$ in the interface of the contract deployed at $\tit{address}$, i.e., $\tit{f}^{\bullet} \in \textit{functions}(\ttbf{s}[\tit{st.address}].\tit{type})$ - the contract's constructor is not a member of this set. 


We abstract the concept of \emph{gas}. In Ethereum, the execution of statements has a cost and each transaction is given a budget, a \emph{gas amount}, that limits the number of statements to be executed. We make a coarse approximation as far as gas is concerned similar to other frameworks~\cite{Kalra18}. Since we are concerned with reachability properties, we generally assume that there is an infinite amount of gas such that computations can run unencumbered by the amount of gas left, and errors can be reached. In Solidity, \tbf{Transfer} or \tbf{Send} constructs trigger the execution of the fallback function of the $destination$ address, if the address hosts a smart contract. However, only a small amount of gas - which does not allow for changes to the execution state - is passed along as a budget for this execution. Hence, we abstract away the existence of this execution altogether. We capture, however, the possibility of a non-deterministic failure in this execution.

The Solid blockchain also abstracts away Ethereum's use of a cryptographic hash function to compute storage locations and new contract addresses. Instead of explicitly computing this function, we see it as an abstract \emph{injective} function. Hence, we capture its usage by assuming that computed storage locations for different reference-type elements are unique and new addresses are fresh. Our function \tit{init_s} ensures that different elements are stored in different reference cells in storage, whereas $\ttbf{s}[\tit{addr}] = \tit{Unused}$ denotes that address $\tit{addr}$ is fresh. As practical hash functions are not injective -- albeit highly improbable, two different inputs can produce the same output: a collision -- in Solidity, two different storage elements could end-up being stored in the same location or a new contract could be deployed in an already active address. However, we specify the Solid blockchain in a way these cases are impossible.

Finally, unlike Solidity/Ethereum, our formalisation eagerly reports errors. Hence, it does not allow, for instance, a function to recover from an error by reverting after the error has happened. We plan to propose an alternative formalisation (and Boogie encoding) where errors are \emph{lazily} reported, more faithfully representing Solidity's behaviour. A far-from-convenient stop-gap solution consists of manually modifying Solidity programs to check for errors before they happen.

These abstractions cause the behaviour of Solid/its blockchain to over-approximate that of Solidity/Ethereum and, therefore, 
it can reach states that are unreachable in Ethereum.


\section{Bounded model checking Solidity with Solidifier}
\label{sec:bmc}

Solidifier is a bounded model checker for Solidity that is based on an encoding of Solid, and its blockchain's behaviour, as a Boogie program, which is later checked by Corral.

\subsection{Solid to Boogie}

Boogie~\cite{Leino08} is a simple but powerful formal (intermediate verification) language that is used to formalise the behaviour of programming languages. Fig.~\ref{fig:boogie} presents some Boogie constructs that we use in our encoding. Boogie does not support the declaration of enumerations and records but they can be simply modelled using other constructs. The concrete syntax of Boogie is in bold font, and we use $\big[$ $\big]$, ? and * to group language elements, declare that elements are optional and express a sequence of elements, respectively. Bold-font square brackets are part of the concrete syntax of Boogie and are used in \tbf{[}\tit{Type}\tbf{]} \tit{Type}' to denote a total mapping from \tit{Type} to \tit{Type}'.

\begin{figure}
	\begin{grammar} \footnotesize
		<Prog> ::= $\big[$<VarDec> | <Proc>$\big]$*
		
		<NameType> ::= <Name> \ttbf{:} <Type>
		
		<TypeDef> ::= \ttbf{enum} <Name> = \ttbf{(}<Name>$\big[$\ttbf{,} <Name>$\big]$*\ttbf{)} \alt \ttbf{record} <Name> = \ttbf{(}<NameType> $\big[$, <NameType> $\big]$*\ttbf{)} 
		
		<VarDec> ::= \ttbf{var} <NameType> 
		
		<Proc> ::= \ttbf{procedure} <Name> \ttbf{(}<NameType>$\big[$\ttbf{,}<NameType>$\big]$*\ttbf{)} \\ \big[\ttbf{returns (}<NameType>\big[\ttbf{,}<NameType>\big]*\ttbf{)}\big]? \{<VarDec>* <Stmt>*\}
		
		<Stmt> ::=  \ttbf{while} (<Expr>) \{<Stmt>*\} \alt \ttbf{if} (<Expr>) \{<Stmt>*\} \big[\tbf{else} \{<Stmt>*\}\big]? \alt <Expr> \ttbf{:=} <Expr> | \tbf{call} <Expr>* \ttbf{:=} <Name>(<Expr>*) \alt \ttbf{assume} <Expr> | \ttbf{assert} <Expr> | \ttbf{havoc} <Expr> | \\tbf{return}
		
		<Type> ::= \ttbf{int} | \ttbf{bool} | \ttbf{[}<Type>\tbf{]} <Type> | \ttbf{enum}(<Name>) \alt \ttbf{record}(<Name>)
		
	\end{grammar}
	\caption{Grammar for a subset of Boogie}
	\label{fig:boogie}
\end{figure}


A Boogie program declares some global variables and procedures. The procedures declare some local variables and a list of statements describing its behaviour. Boogie not only supports traditional commands such as while-loop, if-then-else, assignment and procedure call, but it also supports specification-based statements. The \boogie{{assume}} command makes the program fail silently, \boogie{{assert}} generates an error if its condition is not met, and \boogie{{havoc}} non-deterministically assigns a new value to its expression\footnote{Boogie only allows \boogie{{havoc}} to act on a variable but it is fairly straightforward to model how it could be applied to a ``left-value" expression.}.  As expressions, Boogie has left-value expressions: identifier, mapping access, record access; integer arithmetic  and boolean expressions; quantified expressions; etc. 

We encode a Solid program $\mathcal{S}$ into a Boogie program $\mathcal{B}$, capturing the behaviour of its smart contracts. Program $\mathcal{B}$ represents the execution states via global variables and smart-contract functions are translated in procedures. It uses the following global variables to represent the execution state.
\begin{lstlisting}[numbers=none,language=Boogie, mathescape=true]
record AddressCell = (type : enum(AddressType), balance: UInt, members : [enum(MemberIds)] Ref, storage : [Ref] RefCell);
record RefCell = (type : enum(RefTypes), value : record(RefValues));
record BType = (time : UInt); 
record TxType = (origin : Address); 
record MsgType = (value : UInt);

var s : [Address] AddressCell;
var b : BType;
var tx : TxType;
var m : [Ref] RefCell;
\end{lstlisting}

Each variable captures its corresponding element in the execution state. We use \boogie{{Ref}}, \boogie{{Address}}, \boogie{{UInt}}, and \boogie{{Int}} to represent types \tit{Ref}, \tit{Address}, \tit{UInt} and \tit{Int}, respectively\footnote{These are implemented using Boogie's \ttbf{int} and \ttbf{where} clauses when appropriate. For a given variable, a \tbf{where} clauses specifies a condition that must be met every time this variable is non-deterministically assigned, i.e. either when initialised or when \ttbf{havoc}ed.}. Let $C_1, \ldots, C_n$ be the names of contracts declared in $\mathcal{S}$, \boogie{{enum AddressType}} has values \boogie{{None}}, \boogie{{SimpleAddress}}, \boogie{{$C_1$}}, $\ldots$,\boogie{{$C_n$}}. The enumeration \boogie{{enum MemberIds}} lists the identifiers of member variables accross all contracts in $\mathcal{S}$. The enumeration \boogie{{enum RefTypes}} lists unique tags for all Solid types in $\mathcal{S}$ whose values can be stored in a reference cell. Finally, let $t_1,\ldots,t_m$ be all the Solid types in $\mathcal{S}$ whose values can be stored in a reference cell, \boogie{{record RefValue}} captures the disjoint-union type of all $\tit{RefType}(t_1)\ldots,\tit{RefType}(t_m)$.

The behaviour of Solid statements is captured in Boogie as follows. While and if-then-else constructs are trivially encoded using their Boogie counterparts. The \tit{update} function is encoded using Boogie's assignment statement; it updates either some location in \boogie{{s[this].storage}} or \boogie{{m}}, or a local variable. Function \tit{set_bal} is simply encoded by a block of Boogie code that updates the balances of the concerned addresses. The semantic command \ttbf{fail} is mapped to a failing \boogie{{assume}} whereas \ttbf{error} is to a failing \boogie{{assert}}. The non-deterministic behaviour of \tbf{Transfer} and \tbf{Send} is captured by a non-deterministic if-then-else Boogie construct, namely, where the condition is given by the non-deterministic boolean value \ttbf{*}. These functions' encodings can be used to implement in Boogie all simple statements in Table~\ref{tab:simple-statements} except for \tbf{AllocMemory}. We use auxiliary functions \tit{Allocation} and \tit{Initialisation} to generate the Boogie code to capture \tbf{AllocMemory}.


The Boogie code denoted by $\tit{Allocation}(\tit{refMap},\tit{refExp},T)$ allocates at reference $\tit{refExp}$ in reference mapping $\tit{refMap}$ a fresh tree of reference cells capturing the structure of Solid type $T$. It is given by:
\begin{lstlisting}[numbers=none,language=Boogie,mathescape=true]
assume m[refExp].type == None; 
m[refExp].type := $T$;
$\tit{Allocate}(\tit{refMap},\tit{refExp},T,\tit{vars})$
\end{lstlisting}
The meta-function $\tit{Allocate}(\tit{refMap},\tit{refExp},T,\tit{vars})$, where \tit{vars} is an auxiliary parameter capturing a list of variable declarations, creates the following Boogie code block. If $T$ is a value type or a composite type with value-type elements, $Allocate(\tit{refMap}, \tit{refExp}, T. \tit{vars}) = $ \boogie{{assume true}}. For $T$ a struct with reference-type members $m_1 : T_1, \ldots, m_n : T_n$:
\begin{lstlisting}[numbers=none,language=Boogie,mathescape=true]
$allocate(\tit{refMap},\tit{refMap}[\tit{\tit{refExp}}].value.m_1,T_1, \tit{vars})$
$\ldots$
$allocate(\tit{refMap}, \tit{refMap}[\tit{refExp}].value.m_n,T_n, \tit{vars})$
$Allocate(\tit{refMap}, \tit{refMap}[\tit{refExp}].value.m_1,T_1, \tit{vars})$ 
$\ldots$
$Allocate(\tit{refMap}, \tit{refMap}[\tit{refExp}].value.m_n,T_n, \tit{vars})$
\end{lstlisting}
For $T$ a Solid array with elements of type $T'$:
\begin{lstlisting}[numbers=none,language=Boogie,mathescape=true]
$allocate(\tit{refMap}, \tit{refMap}[\tit{refExp}].value.data[v],T',vars\cat \Seq{v : Int})$ 
$Allocate(\tit{refMap}, \tit{refMap}[\tit{refExp}].value.data[v],T',vars\cat \Seq{v : Int})$
\end{lstlisting}
For $T$ a Solid mapping from type $T'$ to type $T''$:
\begin{lstlisting}[numbers=none,language=Boogie,mathescape=true]
$allocate(\tit{refMap}, \tit{refMap}[\tit{refExp}].value[v],T'',$
					$\tit{vars}\cat\Seq{v : RefType(T')})$ 
$Allocate(\tit{refMap}, \tit{refMap}[\tit{refExp}].value[v],T'',$
					$\tit{vars}\cat\Seq{v : RefType(T')})$ 
\end{lstlisting} 
For $vars \neq \emptySeq$, $allocate(\tit{refMap}, \tit{refExp}, T, \tit{vars})$ corresponds to the following block of Boogie code; $\tit{vars}'$ are new variables mirroring $\tit{vars}$ and $\tit{refExp}'$ is the result of replacing variables in $\tit{vars}$ by their counterparts in $\tit{vars}'$ for $\tit{refExp}$, and $\tit{diff}(\Seq{v_1,\ldots, v_n},\Seq{v'_1,\ldots, v'_n}) = v_1 \neq v'_1 \lor \ldots \lor v_n \neq v'_n$.
\begin{lstlisting}[numbers=none,language=Boogie,mathescape=true]
assume $\forall vars \bullet \tit{refMap}[\tit{refExp}].type = \tit{None}$;
assume $\forall vars, vars' \bullet \tit{diff}(vars,vars') \Rightarrow \tit{refExp} \neq \tit{refExp}'$;
$\tit{preRefMap}$ := $\tit{refMap}$;  
havoc $\tit{refMap}$; 
assume $\forall r : Ref \bullet \tit{preRefMap}[r].type \neq \tit{None}$ 
					$\Rightarrow \tit{preRefMap}[r] = \tit{refMap}[r]$;
assume $\forall vars \bullet \tit{refMap}[\tit{refExp}].type = T$;
\end{lstlisting}
If $vars = \emptySeq$, however, it then corresponds to:
\begin{lstlisting}[numbers=none,language=Boogie,mathescape=true]
assume $\tit{refMap}[\tit{refExp}].type = None$; 
$\tit{refMap}[\tit{refExp}].type$ := $T$;
\end{lstlisting}

To illustrate this construction, we present the Boogie code block created by $\tit{Allocation}(\boogie{{m}}, \boogie{{refExp}},\solidity{{S}})$ where \solidity{|struct S {uint a; uint[] b; uint[][] c}|}:
\begin{lstlisting}[numbers=none,language=Boogie,mathescape=true]
assume m[refExp].type == None;
m[refExp].type := S;
assume m[m[refExp].value.b].type == None;
m[m[refExp].value.b].type := $\tit{uint[]}$;
assume $\forall$ v : int $\bullet$ 
	m[m[m[refExp].value.c].value.data[v]].type 
		== None;
assume $\forall$ v : int, v' : int $\bullet$ v != v' $\Rightarrow$
	m[m[m[refExp].value.c].value.data[v]].value != 
		m[m[m[refExp].value.c].value.data[v]].value;
preM := m;  
havoc m; 
assume $\forall$ r : Ref $\bullet$ 
	preM[r].type != None $\Rightarrow$ preM[r] == m[r];
assume $\forall$ v : int $\bullet$ 
	m[m[m[refExp].value.c].value.data[v]].type 
		== $\tit{uint[]}$;
\end{lstlisting}

As Solid types cannot be recursive, the computation of Boogie code with $\tit{Allocation}$ is guaranteed to finish. Note that no two different reference-type elements can point to the same reference cell. We use $\tit{Initialisation}(\tit{refMap},\tit{refExp},T)$ to denote a sequence of assume statements used to initialise the (reference-cell tree rooted at) reference $\tit{refExp}$ in reference mapping $\tit{refMap}$ to the default value of Solid type $T$. It works similarly to its $Allocation$ counterpart. Executing $\tit{Allocation}(\boogie{{m}},\tit{ref},T)$ followed by $\tit{Initialisation}(\boogie{{m}},\tit{ref},T)$ causes an update from $\boogie{{m}}$ to $\boogie{{m}}' = \tit{alloc}(\boogie{{m}},\tit{ref},T)$, where $\tit{ref}$ is a reference that satisfies $\tit{unalloc}(\tit{refMap},\tit{ref},T)$, as per \tbf{AllocMemory} rule.

The Solid function-call statements in Table~\ref{tab:function-call-statements} are encoded as Boogie procedure calls. A contract call is encoded as the call to a procedure corresponding to the contract function called. Solid's \tbf{Call} is encoded by procedure \boogie{{callP}} and the function \tit{init_s} used to implement \tbf{CreateContract} is implemented by the Boogie code block \tit{DeployContract}; both are defined later. Boogie's argument passing mechanism captures the effects of function \tit{init_l}. As for Solid expressions, they are encoded using their corresponding Boogie counterpart. Arithmetic expressions are encoded using Boogie's operations on mathematical integers with the addition of wrap-around behaviour to capture overflow or underflow for Solidity's bounded integers. This choice is based on the findings in~\cite{Hajdu19}.

Our Boogie encoding relies on \emph{lazy contract deployment}. A function might call a yet uninitialised address. In this case, before the function is executed, the appropriate contract is deployed at this address. For contract $C$ with member variables $m_1 : T_1, \ldots, m_n : T_n$, the lazy deployment is carried out by block $\tit{LazyContractDeployment}(C)$:
\begin{lstlisting}[numbers=none,language=Boogie,mathescape=true]
if (s[this].type == Unused) {
	s[this].type := C;
	$\tit{Allocation}$(s[this].storage,
		s[this].members[$m_1$],$T_1$);
	$\ldots$
	$\tit{Allocation}$(s[this].storage,
		s[this].members[$m_n$],$T_n$);
}
assume s[this].type == C;
\end{lstlisting}

Functions also have a condition that ensures that memory references are well typed. For any reference-type parameter $p$ with type $T$ in Solid function $f$ of contract $C$, block $\tit{MemoryReferenceParametersWellTyped}(C, f)$ creates \boogie{{assume m[$p$].type == $T$}}.

Program $\mathcal{B}$ uses the following procedure to capture the behaviour of Solid function $f$ of contract $C$ with name $\tit{funcName}$, input parameters names $ip_1,\ldots,ip_{ipn}$ with corresponding types $ipT_1,\ldots, ipT_n$, out parameters names $op_1,\ldots,op_{opn}$ with types $opT_1,\ldots, opT_{opn}$, and local variables names $lv_1,\ldots,lv_{lvn}$ with types $lvT_1,\ldots,lvT_{lvn}$.

\begin{lstlisting}[numbers=none,language=Boogie,mathescape=true]
procedure C_funcName($ip_1$ : $LocalType(ipT_1)$,
										$\dots$, 
										$ip_{ipn}$ : $LocalType(ipT_{ipn})$)
	returns ($op_1$ : $LocalType(opT_1)$,
					$\dots$, 
					$op_{ipn}$ : $LocalType(ipT_{ipn})$) 
{
	$\tit{AuxiliaryVariables}$
	var $lv_1$ : $LocalType(T_1)$;
	$\ldots$
	var $lv_{lvn}$ : $LocalType(T_{lvn})$
	$\tit{LazyContractDeployment}(C)$
	$\tit{MemoryReferenceParametersWellTyped}(f)$
	$\tit{Statements}(f.body)$
}
\end{lstlisting}

While $\tit{AuxiliaryVariables}$ declares, for instance, variables to capture pre-memory and pre-storage for allocation purposes, $\tit{Statements}(f.body)$ translates $f.body$ into Boogie. The constructor uses $\tit{DeployContract}(C)$ defined in the following instead of $\tit{LazyContractDeployment}(C)$.
\begin{lstlisting}[numbers=none,language=Boogie,mathescape=true]
s[this].type := C;
s[this].balance := 0;
$\tit{Allocation}$(s[this].storage,
	s[this].members[$m_1$],$T_1$);
$\ldots$
$\tit{Allocation}$(s[this].storage,
	s[this].members[$m_n$],$T_n$);
$\tit{Initialisation}$(s[this].storage,
	s[this].members[$m_1$],$T_1$);
$\ldots$
$\tit{Initialisation}$(s[this].storage,
	s[this].members[$m_n$],$T_n$);
\end{lstlisting}

Note that while $\tit{LazyContractDeployment}(C)$ deploys an instance of $C$ where the values of member variables and its balance are non-deterministically initialised, $\tit{DeployContract}(C)$ creates an instance with their default values.

\subsection{Verification harnesses}

We propose two \emph{verification harnesses}, a contract and a function harness, to drive the exploration of the blockchain's behaviour searching for errors. They define a \boogie{{main}} procedure, and a \boogie{{callP}} procedure capturing Solid's \tbf{Call} construct.

\subsubsection{Contract harness}

For a contract of type $C$, the \boogie{{main}} procedure of the contract harness creates an instance of $C$ at the address given by global variable \boogie{{main_contract}}, by calling its constructor, and executes an arbitrary sequence of interface functions. This procedure ensures that the initial non-deterministically-created reference cells of \boogie{{m}} are well typed and that initially \boogie{{s}} records each address as either unitialised or a simple address. For each new interface function call (the constructor is not an interface function), it non-deterministically initialises \boogie{{b}}, \boogie{{tx}} and the arguments for the function being called. For each new call, it also havocs the storage and balance of all addresses except for \boogie{{main_contract}}: basic values can change but references and reference-cell types are fixed.

The \tbf{Call} construct is treated as an external/unknown function call. The procedure \boogie{{callP}} can execute any (finite) sequence of interface functions of $C$ on \boogie{{main_contract}}. Unlike our formalisation, we do not non-deterministically chose a function in the interface of the contract being called.

Since this harness precisely (up to abstractions) deploys and executes the contract $C$ at address \boogie{{main_contract}}, the interface functions calls are executed from valid/reachable states of $C$, namely, states satisfying $C$'s invariants.

\subsubsection{Function harness}

Our function harness executes a single call to a chosen function $f$ of contract $C$. It analyses the behaviour of $f$ when execution from a non-deterministically initialised state of $C$. The \boogie{{main}} procedure simply calls the procedure corresponding to $f$, whereas \boogie{{callP}} havocs (in a controlled way) storage and balance of all addresses to denote that some arbitrary execution took place.

Unlike our contract harness, the function $f$ might execute from a state that is not reachable by a well-behaved instance of contract $C$, thus possibly reaching actually unreachable errors. However, this harness can find deep errors that might not be reachable by our contract harness, given a fixed bound. While the contract harness examine bounded executions starting in the contract's initial state, the function harness examines bounded executions starting from any state.

\subsection{Solidifier}

Solidifier carries out a bounded verification of Solidity programs looking for failing assertions. It translates a Solidity program into a corresponding Solid program, which is later encoded into Boogie and checked with Corral. It runs the Solidity compiler\footnote{https://github.com/ethereum/solidity} as an auxiliary tool to check if the input Solidity program is valid and to generate its typed abstract syntax tree, later parsed by Solidifier. The user of Solidifier chooses between a contract or function harness. Solidifier also interprets Corral results and information back into Solidity. Aside from the verification primitives declared, functions prefixed with \tit{CexPrint_} and with a single basic-type parameter can be declared in the \boogie{{Verification}} library. When called, they inform Solidifier to print the argument's value at that point in a counterexample, if one reaching this call exists.
Solidifier might output: a counterexample leading to a failing assertion, that no assertion failing assertion can be ever reached, that no failling assertion could be reached for the predetermined bound, or it might fail to produce any output if the underlying solver is unable to solve the constraints generated by Corral\footnote{The use of undecidable theories such as non-linear arithmetic, which we use to capture the modulo and multiplication operation, might lead the underlying solver to an inconclusive result.}.

Solidifier is incomplete by nature as it only explores the behaviour of a contract up to a given bound and it is sound with respect to our verification abstractions on the behaviour of Solid/its blockchain but it is unsound with respect to Solidity/Ethereum. Assuming an infinite amount of gas for computations or that the hash function used by Solidity is collision-free as we do, for instance, makes Solidifier unsound with respect to Solidity/Ethereum: some violations reported by our tool might not actually happen in Solidity/Ethereum. However, these abstractions, also made by other tools~\cite{Wang18,Hajdu19}, are useful for verification as they remove some often unnecessary details from our encoding, leading to better performance and accuracy. 
We make a considerable effort to explain the abstractions we make with respect to Solidity/Ethereum. Thus, Solidifier should be used with this caveat in mind. 

\subsection{Evaluation}

\begin{table}[t]
	\centering
	\caption{Results for Azure workbench examples.}
	\label{tab:results-original-azure}
		\resizebox{.8\textwidth}{!}{%
	\begin{tabular}{|c|c|c|c|c|}
		\hline
		Example & Solidifier & solc-verify & VeriSol & Mythril \\
		\hline
		AssetTransfer & 2.24 ($\times$) & 0.74 ($\times$) & 2.68 ($\times$) & 79.22 ($\times$)  \\
		\hline
		BasicProvenance & 0.91 ($\times$) & 0.46 ($\times$) & 1.74 ($\times$) & 10.18 ($\times$) \\
		\hline
		ItemListingBazaar & 3.76 ($\times$) & 0.53 ($\times$) & 3.93 ($\times$)  & 106.18 ($\times$) \\
		\hline
		DigitalLocker & 1.72 ($\times$) & 0.56 ($\times$) & 1.93 ($\times$) & 36.51 ($\times$) \\
		\hline
		DefectiveComponentC & 6.29 ($\times$) & 0.49 ($\times$) & 5.11 ($\times$) & 1.97 (\tit{bd}) \\
		\hline
		FrequentFlyerRC & 0.71 ($\checkmark$) & \cellcolor{lightgray}0.51 ($\bf \times$) & * & * \\
		\hline
		HelloBlockchain & 0.74 ($\times$) & 0.46 ($\times$) & 1.58 ($\times$) & 5.81 ($\times$) \\
		\hline
		PingPongGame & 1.91 ($\times$) & 0.52 ($\times$) & 3.41 ($\times$) & 184.28 ($\times$) \\
		\hline
		RefrigeratedTrans & 1.01 ($\times$) & 0.50 ($\times$) & 1.96($\times$) & 86.67 ($\times$) \\
		\hline
		RefrigeratedTransWithTime & 1.44 ($\times$) & 0.49 ($\times$) & 1.96 ($\times$) & 112.04 ($\times$) \\
		\hline
		RoomThermostat & 0.66 ($\checkmark$) & 0.46 ($\checkmark$) & \cellcolor{lightgray} 1.62 ($\times$) & 11.70 (\tit{ot}) \\
		\hline
		SimpleMarketplace & 0.93 ($\times$) & 0.48 ($\times$) & 1.73 ($\times$) & 10.68 ($\times$)  \\
		\hline
	\end{tabular}}
\end{table}

\begin{table}[b]
	\centering
	\caption{Results for Azure workbench fixed examples.}
	\label{tab:results-fixed-azure}
		\resizebox{.8\textwidth}{!}{%
		\begin{tabular}{|c|c|c|c|c|}
			\hline
			Example & Solidifier & solc-verify & VeriSol & Mythril  \\
			\hline
			AssetTransfer & 1.43 ($\checkmark$) & 0.57 ($\checkmark$) & 2.10 (\tit{bd}) & 1.20 (\tit{bd})  \\
			\hline
			BasicProvenance & 0.63 ($\checkmark$) & \cellcolor{lightgray} 0.48 ($\times$) & * & 10.28 (\tit{bd})  \\
			\hline
			ItemListingBazaar & 2.26 ($\checkmark$) & \cellcolor{lightgray} 0.54 ($\times$) & * & 111.13 (\tit{ot})  \\
			\hline
			DigitalLocker & 1.02($\checkmark$) & \cellcolor{lightgray} 0.56 ($\times$) & * & 35.27 (\tit{bd}) \\
			\hline
			DefectiveComponentC & 0.70 ($\checkmark$)& \cellcolor{lightgray}  0.49 ($\times$) & * & 2.05 (\tit{bd}) \\
			\hline
			HelloBlockchain & 0.66 ($\checkmark$) & 0.46 ($\checkmark$) & 1.62 (\tit{bd}) & 5.53 (\tit{bd})\\
			\hline
			PingPongGame & * &\cellcolor{lightgray} 0.52 ($\times$) & 30.38 (\tit{bd}) & 189.63 (\tit{bd}) \\
			\hline
			RefrigeratedTrans & 0.88 ($\checkmark$) & \cellcolor{lightgray} 0.50 ($\times$) & * & 87.03 (\tit{bd}) \\
			\hline
			RefrigeratedTransWithTime & 0.89 ($\checkmark$) & \cellcolor{lightgray} 0.51 ($\times$) & *  & 111.93 (\tit{bd}) \\
			\hline
			SimpleMarketplace & 0.64 ($\checkmark$) & 0.48 ($\checkmark$) & * & 11.11 (\tit{bd}) \\
			\hline
	\end{tabular}}
\end{table}

We illustrate the capabilities of Solidifier by comparing it against solc-verify~\cite{Hajdu19,Hajdu20}, VeriSol~\cite{Wang18} and Mythril~\cite{Mythril} in the analysis of 23 contracts. We conducted our evaluation in a Docker container running on a MacBookPro with Intel(R) Core(TM) i7-8559U CPU @ 2.70GHz and 16GB of RAM, and with Docker Engine 18.09.1. Instructions to build this container, the Solidifier binary, and the examples used in this evaluation can be found in \cite{Solidifier}.

Tables~\ref{tab:results-original-azure} and~\ref{tab:results-fixed-azure} present the results of analysing the Solidity contracts of the Azure blockchain workbench~\cite{AzureBlockchainWorkbench}. These smart contract samples are accompanied by an informal textual specification describing what high-level transitions can happen, which we captured with assertions. We fixed the contracts that did not meet their corresponding specification. While Table~\ref{tab:results-original-azure} presents the result for the original contracts, Table~\ref{tab:results-fixed-azure} presents the results for the fixed ones. For each program, we give the time taken in seconds to analyse the contract with the outcome in parenthesis: $\times$ means that a violation was found, $\checkmark$ that the properties have been proved, \tit{bd} represents that no violation has been found in the portion of behaviour analysed (within a transition bound, for instance), \tit{ot} that some violation for an unspecified property has been found (it only applies for Mythril as it does a comprehensive analysis of ``dangerous" code/execution patterns and not only assertion checking), \tit{Error} reports a tool's internal error, * represents a timeout in the analysis of the contract, - denotes that fixing was not needed. We set a 300 seconds timeout for analysing each program. The cells with a grey background present a wrong output: they either state that a violation exists when it does not or that a property holds when it does not. We refer to precision as relating to the number of wrong outputs; the more precise the tool is the fewer wrong outputs it provides. Solidifier uses Corral with a fixed recursion/unwinding bound of 128; this bound is also set for VeriSol.

The properties that we specify for these contracts are mostly local; they restrict pre- and post-states of functions calls. Corral's lazy-inlining over-approximation is usually sufficient to (unconditionally) prove such properties~\cite{Lal12}. For Stater in PingPongGame, however, the property is not local and proving it requires deriving a non-trivial invariant for a recursion.
 
\begin{table}[t]
	\centering
	\caption{Results for other examples.}
	\label{tab:results-others}
	\resizebox{.8\textwidth}{!}{%
	\begin{tabular}{|c|c|c|c|c|}
		\hline
		Example & Solidifier & solc-verify & VeriSol & Mythril   \\
		\hline
		Aliasing & 0.90 ($\checkmark$) & 0.47 ($\checkmark$) & \cellcolor{lightgray} 1.76 ($\times$) & 6.89 (\tit{bd}) \\
		\hline
		StorageDeterministicLayout & 0.83 ($\checkmark$) & \cellcolor{lightgray} 0.48 ($\times$) & 1.79 ($\times$) & * \\
		\hline
		OpenAuction & * &0.49 ($\checkmark$) & 1.98 (\tit{bd}) & 20.10 (\tit{ot}) \\
		\hline
		OpenAuctionWithCall & 2.67 ($\times$) & 0.52 ($\times$) & \tit{Error} & 20.16 (\tit{ot}) \\
		\hline
		Voting & * & \cellcolor{lightgray} 0.57 ($\times$) & \cellcolor{lightgray} 2.10 ($\times$) & * \\
		\hline
		Wallet & 1.72 ($\times$) & \tit{Error} & 1.87 ($\times$) & 25.30 (\tit{bd})\\
		\hline
		WalletNoOverflow & * & \tit{Error} & \cellcolor{lightgray} 1.81 ($\times$) & 27.60 (\tit{bd}) \\
		\hline
		WalletNoOverflowButCall & 1.84 ($\times$) & \tit{Error} & \tit{Error} & 43.80 (\tit{ot})\\
		\hline
		WalletNoOverflowButCallWithLocking & * & \tit{Error} & \tit{Error} & 50.84 (\tit{ot})\\
		\hline
	\end{tabular}}
\end{table}

Table~\ref{tab:results-others} presents other examples with violations arising from more intricate behaviours such as integer overflows or re-entrancy. Examples OpenAuction, OpenAuctionWithCall, and Voting are modified versions of their counterparts in the Solidity documentation~\cite{Solidity}, whereas the other examples were created by us. For OpenAuction(WithCall) we capture the correctness of refunding someone's bid, whereas for Voting we capture the invariant that the number of votes initially available is equal to the sum of votes still to be cast plus votes already cast throughout the election process. For Aliasing and StorageDeterministicLayout, we use our function harness to analyse function $t$. While Aliasing captures properties arising from Solidity's memory model and its different aliasing/deep-copy possibilities, StorageDeterministicLayout tries to capture properties arising from the deterministic organisation of member variables in storage. Finally, all the versions of Wallet, based on Fig.~\ref{fig:wallet}, intend to capture (code and properties of) a typical use of smart contracts to tokenise digital assets.

These results are entirely consistent with how these different tools operate. Mythril is the archetypical traditional tool that uses symbolic execution to explore some portion of a contract's EVM-/low-level behaviour. As such, it should be the most semantically-faithful checker but this faithfulness comes at the expense of speed: it does not report any imprecise result but its analyses take much longer than the others. It should be noted that Mythril checks and reports errors for multiple properties. solc-verify uses the Boogie verifier and its intrinsic over-approximative (assume/guarantee) style of verification. Without user-input annotations (such as contract invariants) to more precisely/narrowly capture the behaviour of a contract, its functions are generally analysed as if they were executed from an arbitrary contract state. Hence, many false errors are reported. VeriSol and Solidifier have similar results in terms of precision, but they differ as far as speed is concerned. We attribute their difference in precision and speed to different Boogie modellings of Solidity as well as to distinct configurations of Corral.

These results seem to suggest that Solidifier fares similarly to other Solidity-to-Boogie-translation-based tools in terms of analysis speed but it is better as far as precision (as in not reporting wrong outcomes) is concerned. Solidifier, VeriSol and solc-verify use different Boogie modellings and support different subsets of Solidity. These aspects should have a direct influence in their precision and speed. Based on these results, it seems that Solidifier captures more faithfully the behaviour of Solidity as compared to these other tools, and specially its memory model, as illustrated by the results Aliasing and StorageDeterministicLayout analyses. The violations that we found were reached with few function calls which seems to suggest that using bounded model checking with a small bound would be enough to find many common errors.

\section{Conclusion}
\label{sec:conclusion}

We have proposed a formalisation of a core subset of Solidity using the Solid language, and a translation of it into Boogie, providing the basis for Solidifier: a bounded model checker that uses Corral to look for semantic-property violations. There are some strong arguments in favour of the approach we propose in comparison to the current state of the art. Firstly, formalising Solidity directly as opposed to EVM bytecode creates a cleaner semantic basis for analysing smart contracts; no need encode low-level operations and let developers bridge the compilation gap from Solidity to bytecode. Secondly, encoding semantic properties, rather than looking for known vulnerable code patterns, allows the designer of a smart contract to precisely specify what is expected from the contract. Their verification might uncover vulnerabilities that are reachable by unknown behaviour patterns or even function errors; neither of which would be found by traditional tools. Thirdly, bounded model checking systematically and symbolically explores all execution paths of a contract up to a certain depth, so it is \emph{guaranteed} to find an error within the given bound. This exploration tends to exercise all code paths of a smart contract but not all execution paths. On the other hand, current ad-hoc symbolic execution techniques for the analysis of smart contracts typically either too coarsely over-approximate the general behaviour of a smart contract instance, leading to inaccurate violations, or they exercise/cover a small portion of the smart contract's code, missing violations. Our evaluation suggests that our approach is more effective at identifying errors for smart contracts. Amongst tools implementing this approach, Solidifier seems to more faithfully capture the semantics of Solidity while remaining as fast as other tools.

A natural question that arises from this work is: what types of contracts are there/can be implemented and what are the semantic properties they should conform to? We plan to conduct a systematic study to address this question.



\section{Acknowledgements}
We thank Ante {\DJ}erek and Liu Han for useful comments on initial drafts of this paper. 

\bibliographystyle{plain}
\bibliography{arxiv}

\end{document}